\documentclass[aps,prl,twocolumn,superscriptaddress,nofootinbib,amsmath,amssymb,letterpaper]{revtex4-2}
\usepackage{amsmath,amssymb,amsthm,mathtools}
\usepackage{graphicx}
\usepackage{microtype}
\usepackage{booktabs}
\usepackage{hyperref}
\usepackage{xcolor}
\definecolor{linkgreen}{RGB}{0,110,60}
\hypersetup{colorlinks=true,linkcolor=linkgreen,citecolor=linkgreen,urlcolor=linkgreen}

\newcommand{\nhat}{\hat{\boldsymbol{n}}}

\usepackage{bm,tikz,pgfplots}
\pgfplotsset{compat=1.18}
\usetikzlibrary{arrows.meta,positioning,calc,shapes.geometric}
\definecolor{navy}{HTML}{17324D}
\definecolor{teal}{HTML}{007F86}
\definecolor{orange}{HTML}{C35A22}
\definecolor{purple}{HTML}{7956A0}
\newcommand{\vv}{\boldsymbol v}
\newcommand{\rr}{\boldsymbol r}
\newcommand{\dd}{\boldsymbol d}

\begin{document}
\title{From Axion Detection to Dark Matter Astronomy}
\author{Jae Hyeok Chang}
\email{jhchang056@snu.ac.kr}
\affiliation{Center for Theoretical Physics, Department of Physics and Astronomy, Seoul National University, Seoul 08826, Korea}
\author{Stefano Profumo}
\email{profumo@ucsc.edu}
\affiliation{Santa Cruz Institute for Particle Physics, 1156 High St., Santa Cruz, CA 95064, USA}
\affiliation{Department of Physics, University of California Santa Cruz, Santa Cruz, CA 95064, USA}

\begin{abstract}
A global haloscope network could turn an axion discovery into astronomy of dark matter substructure and transients. We combine interferometric phase, spatial coherence, and spectroscopy to study the formation and evolution of axion substructure through measurements of flow directions and velocity distributions. For tidal streams, a four-site forecast shows how longitudinal and transverse velocity dispersions can be measured jointly. Interpreted with dynamical and population models, these measurements could test axion production scenarios. Extensions to miniclusters, axion stars, and emitted bursts would broaden the astrophysical reach of laboratory detection.
\end{abstract}
\maketitle

The axion offers a dynamical solution to the strong CP problem~\cite{Peccei:1977hh,Weinberg:1977ma,Wilczek:1977pj} and can account for the cosmological dark matter abundance~\cite{Preskill:1982cy,Abbott:1982af,Dine:1982ah}. Its cosmological production and subsequent gravitational evolution can also generate rich substructure. In scenarios where the Peccei--Quinn symmetry is broken after inflation, this includes miniclusters~\cite{Hogan:1988mp,Kolb:1993zz,Eggemeier:2019khm}, axion stars~\cite{Tkachev:1991ka,Levkov:2018kau,Eggemeier:2019jsu,Gorghetto:2024vnp,Chang:2024fol}, and tidal debris~\cite{Tinyakov:2015cgg,Kavanagh:2020gcy,OHare2024,Visinelli:2026rbm}. The dynamics of compact axion configurations can also produce bursts, including axinovae from collapsing axion stars~\cite{Levkov:2016rkk,Fox:2023xgx,Eby} and bosenova events involving axion clouds around rotating black holes~\cite{Yoshino:2012kn}. These structures and transients offer potential probes of axion production, interactions, and Galactic evolution.

Axion dark matter searches are advancing in sensitivity and mass coverage with cavity haloscopes such as ADMX~\cite{ADMX:2025vom}, CAPP~\cite{CAPP:2024dtx}/DMAG~\cite{Bae:2025bqr}, HAYSTAC~\cite{HAYSTAC2025}, and ORGAN~\cite{ORGAN2025}. Their resonant cavities enhance axion conversion into photons in a magnetic field~\cite{Sikivie:1983ip,Sikivie:2020zpn}. Following a discovery, experiments covering the axion frequency could observe the same narrow band, naturally forming a global network. Interferometry requires simultaneous voltage or complex in-phase and quadrature data with linked timing and frequency references. We illustrate its astronomical reach with tidal streams produced when stellar encounters strip axion miniclusters, small gravitationally bound clumps of dark matter. The debris expands and Galactic orbital shear stretches it into streams~\cite{Tinyakov:2015cgg,Kavanagh:2020gcy,OHare2024,Visinelli:2026rbm}, whose local bulk flow can have a much smaller internal velocity spread.

For cavities with negligible directional response, spectra probe the axion speed distribution~\cite{OHare:2017yze,Foster:2017hbq,Hoskins2016}, but streams with nearly identical spectra can differ in their spread of arrival directions. Directional haloscope designs~\cite{Knirck2018} and correlations among multiple cavities~\cite{ReinaValero2025} can also probe velocity directions. Spatial correlations~\cite{Derevianko:2016vpm} enable dark matter interferometry with geographically separated detectors and Earth's rotation~\cite{Foster:2020fln}. The interferometric phase, visibility amplitude, and line profile constrain a stream's radiant (the mean incoming direction), angular spread, and speed distribution, respectively. Building on network interferometry of dark matter flows~\cite{Foster:2020fln}, we forecast the joint recovery of independent longitudinal and transverse velocity dispersions over a finite stream observation. At leading order in the internal velocities, the single-receiver spectrum is insensitive to the transverse dispersion. We marginalize over unknown station phases and coherence losses to quantify how calibration and network geometry limit measurements of the stream's internal anisotropy.

We consider a single spectrally resolvable stream component whose density and velocity distribution remain approximately constant across the network and during the observation. Calculations predict overlapping minicluster streams with fine spectral structure~\cite{OHare2024,OHare:2025jpr}, while dense components from recent disruptions may be rare~\cite{Visinelli:2026rbm}. Our forecasts quantify the precision on local velocity structure conditional on observing such a component.

We describe this component by a normalized Gaussian velocity distribution $F(\boldsymbol w)$, where $\boldsymbol w$ is an individual axion's velocity relative to Earth's center and $\int d^3w\,F(\boldsymbol w)=1$. Its mean is $\vv=v\hat{\boldsymbol e}$, with bulk speed $v$ and unit flow direction $\hat{\boldsymbol e}$. The radiant $\nhat=-\hat{\boldsymbol e}$ is the mean incoming direction and need not point to the progenitor's current position. The fluctuations $\delta\boldsymbol w=\boldsymbol w-\vv$ have covariance $\boldsymbol\Sigma=\langle\delta\boldsymbol w\,\delta\boldsymbol w^T\rangle$, where brackets denote ensemble averages. Its diagonal elements are the component variances and its off-diagonal elements are their cross covariances. For the benchmark, we additionally assume that the velocity distribution is axisymmetric about the bulk flow, giving
\begin{equation}
 \boldsymbol\Sigma=\sigma_\parallel^2\hat{\boldsymbol e}\hat{\boldsymbol e}^{T}
 +\sigma_\perp^2(\boldsymbol I-\hat{\boldsymbol e}\hat{\boldsymbol e}^{T}).
 \label{eq:tensor}
\end{equation}
Here $\sigma_\parallel$ is the standard deviation along the bulk flow, while $\sigma_\perp$ is the common standard deviation in the two perpendicular directions. For $\sigma_\perp\ll v$, each of the two small angular deviations from the radiant has standard deviation $\sigma_\perp/v$. The assumed symmetry axis follows the bulk flow relative to the observer and need not align with the stream's spatial elongation. When the data support additional parameters, a general velocity ellipsoid can accommodate unequal transverse dispersions and principal axes that are not aligned with the bulk flow.

We use $c=\hbar=1$ in analytic expressions and display units explicitly for numerical values. For particle speed $w=|\boldsymbol w|$, the nonrelativistic signal frequency is $f\simeq f_a(1+w^2/2)$. Here $f_a=m_a/(2\pi)$ is the rest-mass frequency and $m_a$ the axion mass. Detector motion must be included through the relative velocity~\cite{OHare:2017yze,Hoskins2016}. We denote the fitted line centroid by $f_c$ and the frequency offset by $\delta f=f-f_c$. To first order, $\delta f=f_a v\,\delta\boldsymbol w\cdot\hat{\boldsymbol e}$, giving
\begin{equation}
 \sigma_f\simeq f_a v\sigma_\parallel.
 \label{eq:linewidth}
\end{equation}
The linewidth $\sigma_f$ is the standard deviation in frequency and measures the longitudinal spread at leading order. Transverse velocities contribute at higher order in the internal speeds, requiring $\sigma_\perp^2\ll v\sigma_\parallel$. A fitted line centroid absorbs the small transverse contribution to the mean kinetic energy.

The complementary spatial measurement follows from the same velocity distribution~\cite{Derevianko:2016vpm,Foster:2020fln}. Let $\mathcal A_i(t)$ denote the complex axion field at site $i$ and observing time $t$, with the rest-mass carrier removed. The indices $i,j$ label receivers, whose positions $\rr_i(t)$ and $\rr_j(t)$ are defined in a nonrotating frame centered on Earth. Their simultaneous baseline $\dd=\rr_i-\rr_j$ is the displacement vector joining the receivers. The normalized field correlation divides their mutual correlation by the local rms field amplitudes:
\begin{equation}
 \begin{aligned}
 \gamma(\dd)&\equiv
 \frac{\langle\mathcal A_i(t)\mathcal A_j^*(t)\rangle}
 {\sqrt{\langle|\mathcal A_i(t)|^2\rangle\langle|\mathcal A_j(t)|^2\rangle}}\\
 &=\int d^3w\,F(\boldsymbol w)e^{i m_a\boldsymbol w\cdot\dd}.
 \end{aligned}
 \label{eq:gamma_def}
\end{equation}
For independent random plane-wave phases, each velocity contributes the baseline phase $m_a\boldsymbol w\cdot\dd$ weighted by $F$. Evaluating this Gaussian average yields
\begin{equation}
 \begin{aligned}
 \gamma(\dd)&=\exp(i m_a\vv\cdot\dd)\\
 &\quad\times\exp\!\left[-\frac{m_a^2}{2}
 (\sigma_\parallel^2d_\parallel^2+\sigma_\perp^2d_\perp^2)\right],
 \end{aligned}
 \label{eq:visibility}
\end{equation}
where $d=|\dd|$, $d_\parallel=\dd\cdot\hat{\boldsymbol e}$, and $d_\perp^2=d^2-d_\parallel^2$. The exponentials encode the bulk phase and decoherence from velocity spread. The normalization gives $\gamma(0)=1$, and $|\gamma|\leq1$ is the intrinsic visibility for the full signal line, before instrumental losses and noise. For $\sigma_\parallel=\sigma_\perp$, this magnitude depends only on the baseline length and is invariant under Earth's rotation.

Phase, visibility, and spectral shape must be fitted together because they arise from the same stochastic field. Let $S_{ij}(f,t)$ be the complex signal cross-power spectral density between receivers $i$ and $j$, with units of power per unit frequency. Its diagonal $S_{ii}$ is the single-receiver signal spectrum, while off-diagonal elements encode the shared signal's amplitude and relative phase. In the resolved-line limit,
\begin{equation}
 \begin{aligned}
 S_{ij}(f,t)&=\sqrt{P_iP_j}\,g(f)\,E_{ij}\,e^{i\Phi_{ij}(f,t)}\\
 &\quad\times e^{-m_a^2\sigma_\perp^2d_{\perp,ij}^2(t)/2},\\
 \Phi_{ij}(f,t)&=m_a(v+x)d_{\parallel,ij}(t)+\psi_i-\psi_j.
 \end{aligned}
 \label{eq:cross}
\end{equation}
Here $\psi_i$ is the unknown instrumental phase offset of receiver $i$, assumed constant during the observation and across the narrow signal band. Only differences are observable, so we set $\psi_1=0$ and fit the remaining relative phases as nuisance parameters. The baseline $\dd_{ij}(t)=\rr_i(t)-\rr_j(t)$ has signed longitudinal projection $d_{\parallel,ij}=\dd_{ij}\cdot\hat{\boldsymbol e}$ and transverse length $d_{\perp,ij}=(|\dd_{ij}|^2-d_{\parallel,ij}^2)^{1/2}$. The normalized Gaussian line profile $g(f)$ has $\int df\,g(f)=1$, centroid $f_c$, and rms width $f_a v\sigma_\parallel$. The integrated signal power is $P_i=\int df\,S_{ii}$, with $S_{ii}=P_i g(f)$. The variable $x=(f-f_c)/(f_a v)$ is the longitudinal velocity offset selected by frequency to leading order. The spectra in Eq.~\eqref{eq:cross} use a common Earth-centered frequency coordinate at a fixed reference epoch. For each trial flow, station time stamps and ephemerides determine the coherent correction for rotational and orbital Doppler evolution. The phase corrections are referenced to each analysis time $t$, preserving the instantaneous baseline phase $\Phi_{ij}(f,t)$ in Eq.~\eqref{eq:cross}. The same transformation is applied to the complex data and the model covariance, retaining the dependence on the trial flow in the likelihood.

The dimensionless phase $\Phi_{ij}$ is the signal's relative phase modulo $2\pi$, including propagation and instrumental offsets. Residual coherence losses enter through $E_{ii}=1$ and $E_{ij}=\eta_i\eta_j$ for $i\ne j$, where $0\leq\eta_i\leq1$ is a station's coherence efficiency. These losses reduce cross correlations while leaving individual powers unchanged. For $E_{ij}>0$, removing the instrumental phase and coherence losses defines the intrinsic frequency-resolved correlation
\begin{equation}
 \begin{aligned}
 \gamma_{ij}(f,t)&\equiv
 \frac{S_{ij}(f,t)e^{-i(\psi_i-\psi_j)}}
 {E_{ij}\sqrt{S_{ii}(f,t)S_{jj}(f,t)}},\\
 |\gamma_{ij}(f,t)|&=e^{-m_a^2\sigma_\perp^2d_{\perp,ij}^2(t)/2}.
 \end{aligned}
 \label{eq:spectral_visibility}
\end{equation}
Frequency selects the longitudinal velocity offset, leaving transverse velocities to disperse the baseline phase. Earth's rotation changes $d_{\perp,ij}$ and hence the visibility, even for $\sigma_\parallel=\sigma_\perp$. This result holds for any $d_{\parallel,ij}$ and is independent of frequency at this order. Integrating the complex cross spectrum over frequency after instrumental corrections restores the longitudinal damping in Eq.~\eqref{eq:visibility}. At fixed bulk speed and baseline geometry, the linewidth and transverse damping scale with $f_a\sigma_\parallel$ and $f_a\sigma_\perp$, respectively. The bulk phase depends separately on $f_a v$.

We adopt $f_a=10$ GHz ($m_a\simeq41\,\mu\mathrm{eV}$), $v=300\,\mathrm{km/s}$, and $\sigma_\parallel=\sigma_\perp=0.10\,\mathrm{m/s}$ as an illustrative benchmark. In the disruption model of Ref.~\cite{Visinelli:2026rbm}, evaluated at $m_a=50\,\mu\mathrm{eV}$, the median longitudinal dispersion of stripped debris is of order $0.1\,\mathrm{m/s}$ for inner-Galaxy orbits and approaches $0.01\,\mathrm{m/s}$ for orbital semimajor axes $a\gtrsim5\,\mathrm{kpc}$. These model-dependent orbital medians do not specify a typical locally detectable stream. We choose equal fiducial dispersions for illustration but fit them independently. We use approximate city coordinates on a spherical Earth to represent ADMX in Seattle, CAPP/DMAG in Daejeon, HAYSTAC in New Haven, and ORGAN in Perth. The chosen radiant has right ascension $120^\circ$ and declination $+50^\circ$. Adding the Sun's Galactic motion and neglecting Earth's orbital velocity gives a Galactic speed of about $360\,\mathrm{km/s}$, compatible with a bound halo orbit. Restoring Earth's orbital motion makes the exact Galactic speed depend on the observing date. The plots also compare a doubled transverse dispersion, $\sigma_\perp=0.20\,\mathrm{m/s}$.

\begin{figure*}[t]
\centering
\resizebox{\textwidth}{!}{\begin{tikzpicture}
\pgfplotsset{layers/contours/.define layer set={axis background,axis grid,pre main,main,axis ticks,axis lines,axis tick labels,axis descriptions,axis foreground}{/pgfplots/layers/standard}}
\pgfplotsset{every axis/.append style={width=5.35cm,height=5.0cm,scale only axis=false,
tick label style={font=\footnotesize},label style={font=\footnotesize},title style={font=\small,text=navy},
grid=major,grid style={gray!15},legend style={font=\scriptsize,draw=none,fill=white},legend cell align=left}}
\begin{axis}[at={(0,0)},anchor=south west,xmin=-10,xmax=10,ymin=0,ymax=1.5,
title={(a) Same line profile},xlabel={Frequency offset [mHz]},ylabel={Line profile / peak},
xtick={-10,0,10},ytick={0,.5,1},legend style={at={(.5,.98)},anchor=north}]
\addplot[teal,line width=1.5pt] coordinates {(-10,0.01124805) (-9.79798,0.01345942) (-9.59596,0.01604665) (-9.393939,0.01906127) (-9.191919,0.02255944) (-8.989899,0.02660199) (-8.787879,0.03125425) (-8.585859,0.03658586) (-8.383838,0.04267039) (-8.181818,0.04958488) (-7.979798,0.05740914) (-7.777778,0.06622501) (-7.575758,0.07611535) (-7.373737,0.0871629) (-7.171717,0.09944897) (-6.969697,0.113052) (-6.767677,0.1280457) (-6.565657,0.1444979) (-6.363636,0.1624676) (-6.161616,0.1820042) (-5.959596,0.2031446) (-5.757576,0.2259115) (-5.555556,0.2503114) (-5.353535,0.2763326) (-5.151515,0.3039434) (-4.949495,0.3330908) (-4.747475,0.3636987) (-4.545455,0.3956672) (-4.343434,0.4288718) (-4.141414,0.4631634) (-3.939394,0.4983679) (-3.737374,0.5342877) (-3.535354,0.5707022) (-3.333333,0.6073696) (-3.131313,0.6440295) (-2.929293,0.6804053) (-2.727273,0.7162074) (-2.525253,0.751137) (-2.323232,0.7848898) (-2.121212,0.8171606) (-1.919192,0.8476477) (-1.717172,0.8760573) (-1.515152,0.9021087) (-1.313131,0.9255384) (-1.111111,0.9461048) (-0.9090909,0.963592) (-0.7070707,0.9778143) (-0.5050505,0.9886186) (-0.3030303,0.9958877) (-0.1010101,0.9995422) (0.1010101,0.9995422) (0.3030303,0.9958877) (0.5050505,0.9886186) (0.7070707,0.9778143) (0.9090909,0.963592) (1.111111,0.9461048) (1.313131,0.9255384) (1.515152,0.9021087) (1.717172,0.8760573) (1.919192,0.8476477) (2.121212,0.8171606) (2.323232,0.7848898) (2.525253,0.751137) (2.727273,0.7162074) (2.929293,0.6804053) (3.131313,0.6440295) (3.333333,0.6073696) (3.535354,0.5707022) (3.737374,0.5342877) (3.939394,0.4983679) (4.141414,0.4631634) (4.343434,0.4288718) (4.545455,0.3956672) (4.747475,0.3636987) (4.949495,0.3330908) (5.151515,0.3039434) (5.353535,0.2763326) (5.555556,0.2503114) (5.757576,0.2259115) (5.959596,0.2031446) (6.161616,0.1820042) (6.363636,0.1624676) (6.565657,0.1444979) (6.767677,0.1280457) (6.969697,0.113052) (7.171717,0.09944897) (7.373737,0.0871629) (7.575758,0.07611535) (7.777778,0.06622501) (7.979798,0.05740914) (8.181818,0.04958488) (8.383838,0.04267039) (8.585859,0.03658586) (8.787879,0.03125425) (8.989899,0.02660199) (9.191919,0.02255944) (9.393939,0.01906127) (9.59596,0.01604665) (9.79798,0.01345942) (10,0.01124805)};
\addlegendentry{$\sigma_\perp=0.10\,\mathrm{m/s}$}
\addplot[orange,dashed,line width=.9pt] coordinates {(-10,0.01124805) (-9.79798,0.01345942) (-9.59596,0.01604665) (-9.393939,0.01906127) (-9.191919,0.02255944) (-8.989899,0.02660199) (-8.787879,0.03125425) (-8.585859,0.03658586) (-8.383838,0.04267039) (-8.181818,0.04958488) (-7.979798,0.05740914) (-7.777778,0.06622501) (-7.575758,0.07611535) (-7.373737,0.0871629) (-7.171717,0.09944897) (-6.969697,0.113052) (-6.767677,0.1280457) (-6.565657,0.1444979) (-6.363636,0.1624676) (-6.161616,0.1820042) (-5.959596,0.2031446) (-5.757576,0.2259115) (-5.555556,0.2503114) (-5.353535,0.2763326) (-5.151515,0.3039434) (-4.949495,0.3330908) (-4.747475,0.3636987) (-4.545455,0.3956672) (-4.343434,0.4288718) (-4.141414,0.4631634) (-3.939394,0.4983679) (-3.737374,0.5342877) (-3.535354,0.5707022) (-3.333333,0.6073696) (-3.131313,0.6440295) (-2.929293,0.6804053) (-2.727273,0.7162074) (-2.525253,0.751137) (-2.323232,0.7848898) (-2.121212,0.8171606) (-1.919192,0.8476477) (-1.717172,0.8760573) (-1.515152,0.9021087) (-1.313131,0.9255384) (-1.111111,0.9461048) (-0.9090909,0.963592) (-0.7070707,0.9778143) (-0.5050505,0.9886186) (-0.3030303,0.9958877) (-0.1010101,0.9995422) (0.1010101,0.9995422) (0.3030303,0.9958877) (0.5050505,0.9886186) (0.7070707,0.9778143) (0.9090909,0.963592) (1.111111,0.9461048) (1.313131,0.9255384) (1.515152,0.9021087) (1.717172,0.8760573) (1.919192,0.8476477) (2.121212,0.8171606) (2.323232,0.7848898) (2.525253,0.751137) (2.727273,0.7162074) (2.929293,0.6804053) (3.131313,0.6440295) (3.333333,0.6073696) (3.535354,0.5707022) (3.737374,0.5342877) (3.939394,0.4983679) (4.141414,0.4631634) (4.343434,0.4288718) (4.545455,0.3956672) (4.747475,0.3636987) (4.949495,0.3330908) (5.151515,0.3039434) (5.353535,0.2763326) (5.555556,0.2503114) (5.757576,0.2259115) (5.959596,0.2031446) (6.161616,0.1820042) (6.363636,0.1624676) (6.565657,0.1444979) (6.767677,0.1280457) (6.969697,0.113052) (7.171717,0.09944897) (7.373737,0.0871629) (7.575758,0.07611535) (7.777778,0.06622501) (7.979798,0.05740914) (8.181818,0.04958488) (8.383838,0.04267039) (8.585859,0.03658586) (8.787879,0.03125425) (8.989899,0.02660199) (9.191919,0.02255944) (9.393939,0.01906127) (9.59596,0.01604665) (9.79798,0.01345942) (10,0.01124805)};
\addlegendentry{$\sigma_\perp=0.20\,\mathrm{m/s}$}
\end{axis}
\begin{axis}[at={(5.85cm,0)},anchor=south west,xmin=0,xmax=24,ymin=0,ymax=1.04,
title={(b) Different visibilities},xlabel={Greenwich sidereal time [h]},ylabel={Intrinsic visibility $|\gamma_{ij}(f_c,t)|$},
xtick={0,6,12,18,24},ytick={0,.5,1}]
\addplot[orange,dashed,thick] coordinates {(0.125,0.639954242) (0.375,0.645545853) (0.625,0.650286675) (0.875,0.654064635) (1.125,0.656789088) (1.375,0.658393935) (1.625,0.658839948) (1.875,0.658116189) (2.125,0.6562404) (2.375,0.653258367) (2.625,0.649242271) (2.875,0.644288162) (3.125,0.638512684) (3.375,0.632049272) (3.625,0.625044017) (3.875,0.61765144) (4.125,0.610030364) (4.375,0.602340063) (4.625,0.594736837) (4.875,0.587371088) (5.125,0.58038496) (5.375,0.573910539) (5.625,0.568068583) (5.875,0.562967725) (6.125,0.558704057) (6.375,0.555361016) (6.625,0.553009453) (6.875,0.551707815) (7.125,0.551502323) (7.375,0.552427088) (7.625,0.554504101) (7.875,0.55774304) (8.125,0.562140891) (8.375,0.567681356) (8.625,0.574334079) (8.875,0.582053735) (9.125,0.590779026) (9.375,0.600431699) (9.625,0.610915662) (9.875,0.622116353) (10.125,0.633900474) (10.375,0.646116248) (10.625,0.658594326) (10.875,0.671149463) (11.125,0.683583052) (11.375,0.695686554) (11.625,0.70724581) (11.875,0.718046146) (12.125,0.727878126) (12.375,0.736543723) (12.625,0.743862624) (12.875,0.749678352) (13.125,0.753863846) (13.375,0.756326178) (13.625,0.7570101) (13.875,0.755900191) (14.125,0.753021462) (14.375,0.748438373) (14.625,0.742252342) (14.875,0.734597893) (15.125,0.725637707) (15.375,0.715556876) (15.625,0.704556715) (15.875,0.692848453) (16.125,0.680647142) (16.375,0.66816604) (16.625,0.655611682) (16.875,0.643179772) (17.125,0.631051971) (17.375,0.619393568) (17.625,0.608352001) (17.875,0.598056112) (18.125,0.588616029) (18.375,0.580123531) (18.625,0.572652752) (18.875,0.566261098) (19.125,0.560990245) (19.375,0.556867126) (19.625,0.553904814) (19.875,0.55210326) (20.125,0.551449827) (20.375,0.551919632) (20.625,0.553475688) (20.875,0.556068879) (21.125,0.559637825) (21.375,0.564108694) (21.625,0.56939504) (21.875,0.575397767) (22.125,0.582005308) (22.375,0.589094121) (22.625,0.596529593) (22.875,0.604167432) (23.125,0.611855597) (23.375,0.619436796) (23.625,0.626751532) (23.875,0.633641647)};
\addplot[teal,solid,thick] coordinates {(0.125,0.894411203) (0.375,0.896358567) (0.625,0.897999746) (0.875,0.899301187) (1.125,0.90023622) (1.375,0.900785642) (1.625,0.900938158) (1.875,0.900690627) (2.125,0.900048144) (2.375,0.899023916) (2.625,0.897638967) (2.875,0.895921663) (3.125,0.893907091) (3.375,0.891636285) (3.625,0.889155359) (3.875,0.886514543) (4.125,0.883767167) (4.375,0.880968617) (4.625,0.878175284) (4.875,0.875443538) (5.125,0.872828739) (5.375,0.870384303) (5.625,0.868160841) (5.875,0.866205375) (6.125,0.864560633) (6.375,0.863264432) (6.625,0.862349148) (6.875,0.861841264) (7.125,0.861761002) (7.375,0.862122027) (7.625,0.862931238) (7.875,0.864188614) (8.125,0.86588715) (8.375,0.868012857) (8.625,0.870544842) (8.875,0.873455472) (9.125,0.87671062) (9.375,0.880270004) (9.625,0.884087632) (9.875,0.88811234) (10.125,0.892288445) (10.375,0.896556504) (10.625,0.900854176) (10.875,0.905117187) (11.125,0.909280383) (11.375,0.913278855) (11.625,0.917049114) (11.875,0.920530301) (12.125,0.923665382) (12.375,0.926402318) (12.625,0.928695165) (12.875,0.930505067) (13.125,0.931801122) (13.375,0.932561073) (13.625,0.932771823) (13.875,0.932429733) (14.125,0.931540709) (14.375,0.930120061) (14.625,0.928192156) (14.875,0.92578986) (15.125,0.922953796) (15.375,0.919731455) (15.625,0.916176166) (15.875,0.91234598) (16.125,0.908302493) (16.375,0.90410964) (16.625,0.899832492) (16.875,0.895536094) (17.125,0.891284352) (17.375,0.887139002) (17.625,0.883158667) (17.875,0.87939802) (18.125,0.875907049) (18.375,0.872730433) (18.625,0.869907025) (18.875,0.867469435) (19.125,0.865443713) (19.375,0.86384912) (19.625,0.862697988) (19.875,0.861995657) (20.125,0.861740494) (20.375,0.861923974) (20.625,0.86253085) (20.875,0.86353938) (21.125,0.864921644) (21.375,0.866643927) (21.625,0.868667193) (21.875,0.870947633) (22.125,0.873437304) (22.375,0.876084854) (22.625,0.878836323) (22.875,0.881636025) (23.125,0.884427493) (23.375,0.88715448) (23.625,0.889761994) (23.875,0.892197349)};
\node[font=\scriptsize,text=navy,anchor=south] at (axis cs:12,.05) {ADMX and CAPP/DMAG};
\end{axis}
\begin{axis}[at={(11.7cm,0)},anchor=south west,set layers=contours,xmin=.065,xmax=.135,ymin=.05,ymax=.17,
title={(c) Forecast precision},xlabel={$\sigma_\parallel$ [m/s]},ylabel={$\sigma_\perp$ [m/s]},
xtick={.07,.1,.13},xticklabels={0.07,0.10,0.13},ytick={.06,.1,.14},yticklabels={0.06,0.10,0.14}]
\path[fill=gray!18] (axis cs:0.08480135283,.05) rectangle (axis cs:0.1179226471,.17);
\addplot[teal,fill=teal!18,thin] coordinates {(0.0869648835,0.100171236) (0.086991887,0.101843137) (0.0870375992,0.103540026) (0.087101979,0.105259542) (0.0871849683,0.106999158) (0.0872864921,0.108756178) (0.0874064581,0.110527745) (0.0875447566,0.112310833) (0.0877012602,0.114102259) (0.0878758237,0.115898681) (0.0880682838,0.117696606) (0.0882784587,0.119492396) (0.0885061479,0.12128228) (0.0887511315,0.123062355) (0.0890131706,0.124828607) (0.0892920062,0.126576919) (0.089587359,0.128303083) (0.0898989293,0.130002819) (0.0902263964,0.13167179) (0.090569418,0.133305621) (0.0909276303,0.134899915) (0.0913006474,0.136450277) (0.0916880609,0.137952332) (0.0920894394,0.139401748) (0.0925043289,0.140794257) (0.0929322517,0.14212568) (0.0933727066,0.143391946) (0.0938251688,0.14458912) (0.0942890897,0.14571342) (0.0947638965,0.146761244) (0.0952489927,0.147729189) (0.0957437579,0.148614073) (0.0962475478,0.149412956) (0.0967596946,0.150123156) (0.0972795074,0.150742268) (0.0978062723,0.151268178) (0.0983392529,0.151699081) (0.0988776912,0.152033489) (0.0994208078,0.152270242) (0.0999678031,0.152408515) (0.100517858,0.152447828) (0.101070134,0.152388042) (0.101623777,0.152229367) (0.102177914,0.151972355) (0.10273166,0.151617901) (0.103284116,0.151167231) (0.103834368,0.150621902) (0.104381496,0.149983785) (0.104924569,0.149255052) (0.105462649,0.148438168) (0.105994797,0.147535869) (0.106520066,0.146551149) (0.107037513,0.145487235) (0.107546196,0.144347574) (0.108045175,0.143135807) (0.108533519,0.141855749) (0.109010305,0.140511365) (0.109474622,0.13910675) (0.109925572,0.137646104) (0.110362275,0.136133709) (0.110783868,0.134573909) (0.111189513,0.132971084) (0.111578393,0.131329631) (0.11194972,0.129653945) (0.112302735,0.127948394) (0.11263671,0.126217304) (0.112950951,0.124464943) (0.113244802,0.1226955) (0.113517643,0.120913072) (0.113768898,0.119121653) (0.113998033,0.117325116) (0.114204555,0.115527208) (0.114388023,0.113731537) (0.11454804,0.111941565) (0.11468426,0.110160602) (0.114796388,0.108391799) (0.114884178,0.106638147) (0.11494744,0.104902469) (0.114986035,0.103187427) (0.11499988,0.101495514) (0.114988943,0.0998290565) (0.114953248,0.0981902201) (0.114892875,0.0965810072) (0.114807954,0.0950032632) (0.114698671,0.0934586797) (0.114565264,0.0919487992) (0.114408022,0.0904750209) (0.114227287,0.0890386059) (0.114023447,0.0876406839) (0.113796942,0.0862822591) (0.113548256,0.0849642174) (0.113277918,0.0836873332) (0.112986501,0.0824522761) (0.11267462,0.0812596184) (0.112342926,0.0801098419) (0.111992108,0.079003345) (0.111622891,0.0779404499) (0.11123603,0.0769214091) (0.110832311,0.0759464119) (0.110412546,0.0750155917) (0.109977572,0.0741290313) (0.109528248,0.0732867697) (0.109065454,0.0724888074) (0.108590084,0.0717351122) (0.108103049,0.0710256243) (0.10760527,0.0703602616) (0.107097677,0.0697389238) (0.10658121,0.0691614973) (0.106056809,0.0686278589) (0.10552542,0.06813788) (0.104987987,0.0676914297) (0.104445451,0.0672883783) (0.103898751,0.0669286) (0.103348817,0.0666119755) (0.102796573,0.0663383944) (0.102242932,0.0661077573) (0.101688794,0.0659199773) (0.101135048,0.065774982) (0.100582566,0.0656727138) (0.100032207,0.0656131318) (0.0994848103,0.0655962117) (0.0989411967,0.0656219468) (0.0984021687,0.0656903474) (0.0978685079,0.0658014412) (0.0973409751,0.0659552729) (0.0968203091,0.0661519028) (0.0963072264,0.0663914068) (0.0958024209,0.0666738743) (0.0953065629,0.0669994072) (0.0948202995,0.067368118) (0.0943442539,0.0677801273) (0.0938790255,0.0682355621) (0.0934251897,0.0687345525) (0.0929832981,0.0692772293) (0.0925538784,0.0698637204) (0.0921374343,0.070494147) (0.0917344463,0.0711686206) (0.0913453714,0.0718872378) (0.0909706433,0.0726500768) (0.0906106733,0.0734571919) (0.0902658498,0.0743086092) (0.0899365394,0.0752043204) (0.0896230866,0.0761442782) (0.0893258148,0.0771283897) (0.0890450262,0.0781565106) (0.0887810023,0.079228439) (0.0885340045,0.0803439084) (0.0883042742,0.0815025816) (0.0880920333,0.0827040436) (0.0878974846,0.0839477943) (0.0877208122,0.0852332421) (0.0875621815,0.086559696) (0.0874217398,0.0879263593) (0.0872996165,0.0893323223) (0.0871959235,0.0907765553) (0.087110755,0.092257902) (0.0870441881,0.0937750732) (0.0869962829,0.0953266407) (0.0869670824,0.0969110314) (0.0869566128,0.0985265224) (0.0869648835,0.100171236)} -- cycle;
\addplot[teal,fill=teal!55,thin] coordinates {(0.0917469725,0.100105571) (0.091764541,0.101132731) (0.0917942767,0.102168656) (0.0918361454,0.103211778) (0.0918900991,0.104260468) (0.0919560759,0.105313037) (0.0920339998,0.106367742) (0.0921237806,0.107422784) (0.0922253144,0.108476314) (0.0923384828,0.109526435) (0.0924631537,0.110571207) (0.0925991806,0.11160865) (0.0927464028,0.11263675) (0.0929046456,0.113653465) (0.0930737198,0.114656729) (0.0932534221,0.115644457) (0.0934435349,0.116614556) (0.0936438265,0.117564927) (0.0938540507,0.118493473) (0.094073947,0.119398109) (0.0943032411,0.120276766) (0.0945416441,0.121127402) (0.0947888532,0.121948008) (0.0950445516,0.122736615) (0.0953084086,0.123491306) (0.0955800797,0.124210218) (0.0958592066,0.124891558) (0.0961454179,0.125533604) (0.0964383288,0.126134715) (0.0967375415,0.126693339) (0.0970426455,0.12720802) (0.097353218,0.127677407) (0.097668824,0.128100255) (0.0979890171,0.128475435) (0.0983133394,0.128801942) (0.0986413223,0.129078892) (0.0989724871,0.129305536) (0.0993063452,0.129481256) (0.0996423989,0.129605571) (0.0999801421,0.129678143) (0.100319061,0.129698771) (0.100658634,0.129667399) (0.100998334,0.129584114) (0.101337629,0.129449143) (0.10167598,0.129262856) (0.102012846,0.129025761) (0.102347683,0.128738504) (0.102679946,0.128401862) (0.103009086,0.128016745) (0.103334557,0.127584183) (0.103655815,0.127105331) (0.103972314,0.126581454) (0.104283516,0.126013928) (0.104588884,0.125404227) (0.104887891,0.124753924) (0.105180012,0.124064676) (0.105464734,0.123338221) (0.105741552,0.122576367) (0.106009969,0.121780988) (0.106269505,0.120954014) (0.106519688,0.120097421) (0.106760062,0.119213227) (0.106990187,0.11830348) (0.107209638,0.117370252) (0.107418007,0.116415632) (0.107614908,0.115441718) (0.107799969,0.114450608) (0.107972844,0.113444394) (0.108133206,0.112425158) (0.108280749,0.11139496) (0.108415194,0.110355839) (0.108536282,0.109309801) (0.108643783,0.108258818) (0.108737489,0.107204821) (0.10881722,0.106149699) (0.108882822,0.105095289) (0.108934168,0.10404338) (0.108971159,0.102995705) (0.108993723,0.101953939) (0.109001816,0.100919699) (0.108995422,0.0998945399) (0.108974555,0.0988799557) (0.108939254,0.0978773761) (0.108889588,0.0968881672) (0.108825653,0.0959136309) (0.108747572,0.0949550054) (0.108655497,0.094013465) (0.108549605,0.0930901211) (0.108430099,0.092186023) (0.108297209,0.091302159) (0.108151189,0.0904394579) (0.107992316,0.0895987904) (0.107820893,0.0887809708) (0.107637244,0.0879867587) (0.107441714,0.0872168613) (0.107234671,0.0864719351) (0.107016499,0.085752588) (0.106787605,0.0850593817) (0.106548411,0.0843928339) (0.106299356,0.0837534202) (0.106040894,0.0831415767) (0.105773494,0.082557702) (0.105497637,0.0820021594) (0.105213816,0.0814752792) (0.104922537,0.0809773607) (0.104624311,0.0805086741) (0.104319662,0.0800694629) (0.104009117,0.0796599451) (0.103693211,0.0792803157) (0.103372484,0.0789307481) (0.103047479,0.0786113955) (0.102718741,0.0783223927) (0.102386817,0.0780638574) (0.102052253,0.0778358912) (0.101715597,0.0776385812) (0.101377392,0.0774720005) (0.10103818,0.0773362094) (0.1006985,0.077231256) (0.100358885,0.077157177) (0.100019862,0.0771139977) (0.099681954,0.077101733) (0.0993456757,0.0771203869) (0.099011534,0.0771699532) (0.0986800277,0.0772504149) (0.0983516464,0.0773617443) (0.0980268701,0.0775039024) (0.0977061686,0.0776768387) (0.0973900009,0.0778804903) (0.097078815,0.0781147812) (0.0967730471,0.0783796215) (0.0964731216,0.0786749063) (0.0961794504,0.0790005145) (0.0958924328,0.0793563076) (0.095612455,0.0797421283) (0.0953398902,0.080157799) (0.0950750982,0.0806031201) (0.094818425,0.0810778683) (0.094570203,0.0815817949) (0.0943307507,0.0821146237) (0.0941003725,0.0826760492) (0.0938793589,0.0832657347) (0.093667986,0.0838833094) (0.0934665158,0.0845283674) (0.0932751963,0.0852004646) (0.0930942608,0.0858991167) (0.0929239286,0.0866237973) (0.0927644048,0.0873739353) (0.09261588,0.0881489128) (0.0924785308,0.088948063) (0.0923525195,0.0897706682) (0.0922379941,0.0906159576) (0.0921350887,0.0914831052) (0.092043923,0.0923712287) (0.0919646029,0.093279387) (0.0918972199,0.0942065793) (0.0918418518,0.0951517435) (0.0917985622,0.0961137553) (0.0917674008,0.0970914273) (0.0917484032,0.0980835081) (0.0917415911,0.0990886826) (0.0917469725,0.100105571)} -- cycle;
\addplot[only marks,mark=*,mark size=2.4pt,teal] coordinates {(0.1,0.1)};
\end{axis}
\end{tikzpicture}
}\caption{Spectral and spatial measurements at $f_a=10$ GHz and fiducial $\sigma_\parallel=0.10\,\mathrm{m/s}$. (a) The two transverse dispersions give the same leading spectrum. (b) Intrinsic visibility at $f=f_c$ on the Seattle--Daejeon baseline, from Eq.~\eqref{eq:spectral_visibility}. (c) The teal circle marks the fiducial $\sigma_\perp=0.10\,\mathrm{m/s}$. Darker and lighter regions are joint $68\%$ and $95\%$ Fisher forecasts using all four individual spectra and six cross spectra. Computed in log dispersions, they marginalize over other parameters at fixed mass and speed within a local directional solution. The gray band is a single receiver's $68\%$ longitudinal interval, with no transverse constraint at this order.}
\label{fig:dispersion}
\end{figure*}
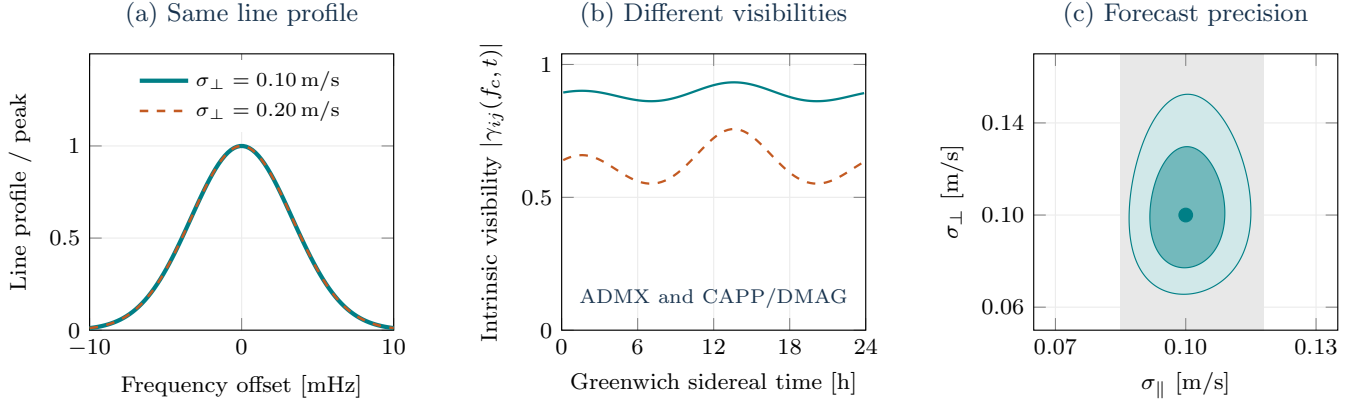

The longitudinal width gives $\sigma_f=3.3$ mHz. For this flow, Earth's rotation produces a relative Doppler shift of up to about $12\,\mathrm{Hz}$ between Seattle and Daejeon, much larger than the linewidth. Earth's orbital acceleration can also shift the common line centroid by up to about $30\,\mathrm{Hz}$ over two sidereal days, depending on the observing date. Increasing $\sigma_\perp$ from $0.10$ to $0.20\,\mathrm{m/s}$ leaves the leading spectrum unchanged but reduces the transverse coherence length $\ell_{\rm coh,\perp}=1/(m_a\sigma_\perp)$ from $1.4\times10^{4}$ to $7.2\times10^{3}\,\mathrm{km}$. For $d_\perp=7.0\times10^{3}\,\mathrm{km}$, the frequency-resolved intrinsic visibility falls from $0.89$ to $0.62$, reaching $e^{-1/2}$ at $d_\perp=\ell_{\rm coh,\perp}$. The one-coordinate angular spread doubles from $0.069''$ to $0.14''$. These intrinsic widths are distinct from uncertainty in the mean flow direction. Figure~\ref{fig:dispersion}(a) compares the line profiles, while Fig.~\ref{fig:dispersion}(b) shows the frequency-resolved visibility at $f=f_c$ on the Seattle--Daejeon baseline as Earth rotates.

After receiver responses are applied, $S=(S_{ij})$ and $N=(N_{ij})$ are the signal and noise spectral covariance matrices. The element $N_{ij}(f,t)$ is the noise cross-power spectral density between receivers $i,j$, in the same units as $S_{ij}$, and $N_i\equiv N_{ii}$ is the individual noise spectrum. Their sum $C(f,t)=S(f,t)+N(f,t)$ is the spectral covariance of the complex Fourier data at frequency $f$ and time $t$~\cite{Foster:2017hbq,Foster:2020fln}. A single network likelihood combines individual and cross spectra while accounting for correlations between shared baselines. Finite crossings enter through the observation window and time-dependent powers and velocity structure. For a stationary velocity distribution, the normalized time-domain kernel is
\begin{equation}
 K(\Delta\rr,\Delta t)=\int d^3w\,F(\boldsymbol w)
 e^{i m_a(\boldsymbol w\cdot\Delta\rr-w^2\Delta t/2)}.
 \label{eq:kernel}
\end{equation}
Here $\Delta\rr=\rr_i(t)-\rr_j(t')$ and $\Delta t=t-t'$ are the spatial separation and time lag between samples. The kernel has the rest-mass carrier removed and reduces to $\gamma(\dd)$ at equal times. Finite windows filter it, while evolving distributions require a time-dependent generalization.

We illustrate dispersion recovery with a controlled post-discovery benchmark using the four-site geometry and stream parameters specified above. Idealized future receivers operating at $10$ GHz observe the same line continuously for $T=2$ sidereal days, with equal, known, independent white-noise spectra and flat response over the line. Thus $N_{ij}=0$ for $i\ne j$. The stream's density and velocity distribution remain constant within this window. The peak signal-to-noise \emph{spectral-density ratio} is assumed to be $P_i g(f_c)/N_i=0.30$ at every site, independently of current experimental performance. This ratio is not a detection significance. The benchmark neglects solar and terrestrial gravitational focusing, which can alter the local density, flow direction, and velocity covariance~\cite{Kim:2021yyo}. The inferred parameters describe the local distribution at Earth. Relating them to an asymptotic stream requires a model of gravitational focusing. We also neglect corrections of relative order rotational speed over $v$ to the residual line shape. The forecast assumes accurate station registration and ephemeris dechirping, with negligible residual frequency drift. Gaussian fluctuations of the residual relative phase on a baseline, with rms $\sigma_{\Delta\phi}$ during averaging, reduce the cross-spectrum amplitude by $e^{-\sigma_{\Delta\phi}^2/2}$. Keeping any additional unmodeled loss below $1\%$ requires $\sigma_{\Delta\phi}\lesssim0.14$ rad, equivalent to about $2.3$ ps of differential timing jitter at $10$ GHz. This value describes residual phase stability and does not impose an absolute clock accuracy requirement. Constant phase offsets are included in $\psi_i$, while stationary coherence losses are represented by $\eta_i$.

The benchmark coherence length is comparable to Earth's diameter, allowing spatial correlations to probe the transverse width. Early estimates of minicluster disruption include stream crossings lasting $2$--$3$ days~\cite{Tinyakov:2015cgg}. Our exposure $T$ is an observing interval within an encounter, not a prediction of its duration. A crossing of this length does not by itself ensure a constant envelope. The forecast is conditional on the component's mean density and velocity distribution remaining approximately stationary over the full two sidereal days. At the benchmark linewidth, the field coherence time $\tau_c=1/(2\pi\sigma_f)\simeq48\,\mathrm{s}$ describes decorrelation of the wave, not the encounter duration. Many field realizations contribute during $T$. The chosen dispersions, signal powers, and exposure illustrate such a detectable component. Their joint occurrence probability and encounter rate require a population model for widths, densities, and ages.

We collect both log dispersions, four independent log signal powers, the centroid $f_c$, two local radiant coordinates, three relative constant phases, and four log coherence efficiencies in the fit parameter vector $\boldsymbol\theta$. For proper complex Gaussian Fourier modes, the local information matrix is~\cite{Foster:2020fln}
\begin{equation}
 \begin{aligned}
 \mathcal I_{ab}&=\int_0^T dt\int df\\
 &\quad\times\operatorname{Tr}\!\left[C^{-1}C_{,a}C^{-1}C_{,b}\right]
 +\mathcal I^{\rm cal}_{ab}.
 \end{aligned}
 \label{eq:fisher}
\end{equation}
Here $a,b$ label the components $\theta_a,\theta_b$ of $\boldsymbol\theta$, $C_{,a}=\partial C/\partial\theta_a$, and the trace sums over receiver indices $i,j$. The time integral covers the coincident exposure $T$, the frequency integral covers the resolved signal line, and $\mathcal I^{\rm cal}$ is the information supplied by the assumed calibration priors. We evaluate Eq.~\eqref{eq:fisher} in the slowly varying, resolved-line approximation.

For this prospective calibration scenario, we set $\eta_i=0.98$ with independent Gaussian priors on $\ln\eta_i$ of standard deviation $\Delta\ln\eta_i=0.010$. These values are assumed future performance rather than demonstrated station specifications. The priors describe uncertainty in the efficiencies and are distinct from the illustrative $1\%$ budget for additional unmodeled coherence loss. Mass and bulk speed are fixed, conditional on an identified local directional solution. The inverse information matrix $\mathcal I^{-1}$ is the local parameter covariance. Converting its width block from log widths to physical dispersions gives the marginalized local $1\sigma$ uncertainties $\Delta\sigma_\parallel$ and $\Delta\sigma_\perp$,
\begin{equation}
 \begin{aligned}
 \Delta\sigma_\parallel&=0.0057\,\mathrm{m/s},\\
 \Delta\sigma_\perp&=0.017\,\mathrm{m/s}.
 \end{aligned}
 \label{eq:result}
\end{equation}
These correspond to $5.7\%$ and $17\%$, with correlation $0.033$. Including this covariance, the anisotropy ratio is $\sigma_\perp/\sigma_\parallel=1.00\pm0.18$ at local $1\sigma$ for fixed bulk speed. Figure~\ref{fig:dispersion}(c) compares the joint Fisher regions with a single receiver at the same exposure and spectral sensitivity, which gives $\Delta\sigma_\parallel=0.016\,\mathrm{m/s}$ and no transverse constraint at leading order. Allowing $v$ to vary within the same local solution gives $\Delta\sigma_\parallel=0.014\,\mathrm{m/s}$ and $\Delta\sigma_\perp=0.017\,\mathrm{m/s}$. Here $m_a$ remains fixed and $f_c$ is an independent nuisance parameter, so the fit uses no speed information from $f_c-f_a$. Longitudinal precision worsens because the linewidth measures $v\sigma_\parallel$. These are local Fisher forecasts conditional on the assumed axion mass, rather than confidence regions from simulated or observed data. A broader analysis could include $m_a$ in the fit and propagate its uncertainty.

\begin{figure}[t]
\centering
\begin{tikzpicture}
\pgfplotsset{layers/resolution/.define layer set={axis background,axis grid,pre main,main,axis ticks,axis lines,axis tick labels,axis descriptions,axis foreground}{/pgfplots/layers/standard}}
\begin{loglogaxis}[set layers=resolution,axis lines=box,width=\columnwidth,height=5.5cm,xmin=.02,xmax=1,ymin=.05,ymax=8,
xlabel={Transverse dispersion $\sigma_\perp$ [m/s]},ylabel={Fractional uncertainty $\Delta\sigma_\perp/\sigma_\perp$},
tick label style={font=\footnotesize},label style={font=\footnotesize},
xtick={.02,.05,.1,.2,.5,1},xticklabels={0.02,0.05,0.1,0.2,0.5,1},
ytick={.1,1},yticklabels={0.1,1},grid=major,grid style={gray!18},
legend style={font=\scriptsize,draw=none,fill=white,at={(.98,.98)},anchor=north east},legend cell align=left]
\path[fill=gray!10] (axis cs:.02,1) rectangle (axis cs:1,8);
\addplot[gray,dotted,forget plot] coordinates {(0.02,1) (1,1)};
\addplot[thick,teal] coordinates {(0.02,3.26149508) (0.0211666803,2.91600382) (0.0224014176,2.60755479) (0.0237081822,2.33217778) (0.0250911756,2.08632828) (0.0265548446,1.86684182) (0.0281038952,1.67089326) (0.0297433082,1.49596045) (0.0314783547,1.33979171) (0.0333146135,1.20037692) (0.0352579886,1.07592157) (0.0373147285,0.96482372) (0.0394914464,0.865653338) (0.0417951409,0.777133925) (0.0442332192,0.698126063) (0.0468135203,0.627612756) (0.0495443408,0.564686333) (0.052434461,0.508536763) (0.0554931736,0.45844123) (0.058730313,0.41375482) (0.0621562879,0.373902216) (0.0657821136,0.338370288) (0.0696194482,0.30670148) (0.07368063,0.278487911) (0.0779787168,0.253366108) (0.0825275282,0.231012313) (0.0873416901,0.211138293) (0.0924366814,0.193487591) (0.0978288839,0.177832185) (0.103535635,0.163969488) (0.109575284,0.151719646) (0.11596725,0.140923095) (0.122732085,0.131438324) (0.12989154,0.123139793) (0.137468635,0.115915962) (0.145487732,0.109667362) (0.153974616,0.104304673) (0.162956573,0.0997467619) (0.172462483,0.0959186636) (0.182522912,0.0927495766) (0.193170206,0.0901709985) (0.204438599,0.0881152867) (0.216364323,0.0865150298) (0.228985722,0.0853036561) (0.242343378,0.0844175297) (0.25648024,0.0837993338) (0.271441762,0.0834018691) (0.287276049,0.0831908403) (0.304034014,0.0831452251) (0.321769538,0.0832546859) (0.340539646,0.0835149257) (0.36040469,0.0839231541) (0.381428542,0.084476157) (0.403678799,0.0851725605) (0.427227003,0.086019106) (0.452148869,0.0870389266) (0.478524526,0.0882789251) (0.506438782,0.0898139768) (0.535981389,0.0917475419) (0.567247334,0.0942103015) (0.600337147,0.0973595693) (0.635357222,0.101382048) (0.672420158,0.106501343) (0.711645124,0.112990117) (0.753158239,0.121185351) (0.797092982,0.131503914) (0.843590614,0.144454052) (0.892800639,0.160635632) (0.944881283,0.180717152) (1,0.205370953)};
\addlegendentry{$1\%$ priors on station efficiencies}
\addplot[thick,purple,dashed] coordinates {(0.02,4.20630259) (0.0211666803,3.75978911) (0.0224014176,3.36114853) (0.0237081822,3.00524976) (0.0250911756,2.68751184) (0.0265548446,2.40384502) (0.0281038952,2.15059809) (0.0297433082,1.92451141) (0.0314783547,1.72267492) (0.0333146135,1.54249072) (0.0352579886,1.3816396) (0.0373147285,1.23805124) (0.0394914464,1.10987749) (0.0417951409,0.995468662) (0.0442332192,0.893352253) (0.0468135203,0.802214014) (0.0495443408,0.720881039) (0.052434461,0.648306679) (0.0554931736,0.583557079) (0.058730313,0.525799176) (0.0621562879,0.474289996) (0.0657821136,0.428367113) (0.0696194482,0.387440155) (0.07368063,0.350983242) (0.0779787168,0.318528269) (0.0825275282,0.289658937) (0.0873416901,0.264005473) (0.0924366814,0.241239964) (0.0978288839,0.22107225) (0.103535635,0.203246332) (0.109575284,0.187537246) (0.11596725,0.173748367) (0.122732085,0.16170911) (0.12989154,0.151272994) (0.137468635,0.142316016) (0.145487732,0.134735301) (0.153974616,0.128447933) (0.162956573,0.123389854) (0.172462483,0.119514649) (0.182522912,0.116791929) (0.193170206,0.115204939) (0.204438599,0.114746904) (0.216364323,0.11541566) (0.228985722,0.117206438) (0.242343378,0.120103522) (0.25648024,0.124072978) (0.271441762,0.129060298) (0.287276049,0.134997048) (0.304034014,0.141817578) (0.321769538,0.149480662) (0.340539646,0.157986009) (0.36040469,0.167377662) (0.381428542,0.177734399) (0.403678799,0.189153134) (0.427227003,0.201729517) (0.452148869,0.215536304) (0.478524526,0.230603524) (0.506438782,0.246914154) (0.535981389,0.264431791) (0.567247334,0.283162055) (0.600337147,0.303225893) (0.635357222,0.324916458) (0.672420158,0.34873341) (0.711645124,0.375416621) (0.753158239,0.406003355) (0.797092982,0.441906967) (0.843590614,0.484986147) (0.892800639,0.537560639) (0.944881283,0.602326187) (1,0.682117738)};
\addlegendentry{No efficiency priors}
\addplot[only marks,mark=*,mark size=2.7pt,teal,forget plot] coordinates {(0.1,0.172262795)};
\node[font=\scriptsize,anchor=south west,text=navy] at (axis cs:.18,1.05) {Width unresolved};
\end{loglogaxis}
\end{tikzpicture}
\caption{Transverse-width sensitivity with exposure, signal powers, and longitudinal width at their benchmark values. The bulk velocity vector is fixed, while the longitudinal width, powers, centroid, station phases, and coherence efficiencies are marginalized. Solid and dashed curves use independent calibration priors $\Delta\ln\eta_i=0.010$ and no such priors, respectively. The circle marks $\sigma_\perp=0.10\,\mathrm{m/s}$.}
\label{fig:resolution}
\end{figure}
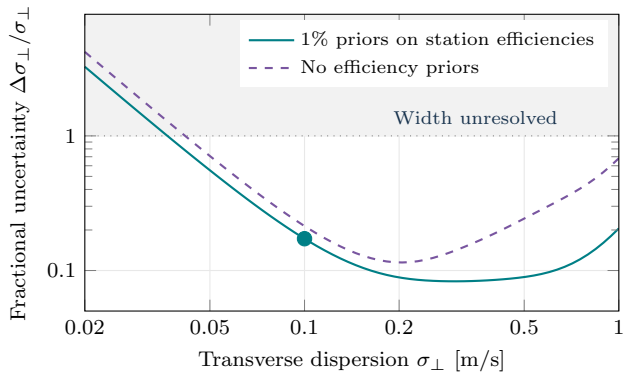

For $m_a d_\perp\sigma_\perp\ll1$, the visibility changes only quadratically with $\sigma_\perp$, and the width becomes difficult to distinguish from zero. At large projected separations the cross signal is suppressed. Figure~\ref{fig:resolution} shows the useful intermediate regime for the same network and signal powers. In the unresolved regime one should report an upper limit using a nonnegative width parameter, rather than extrapolate a local Gaussian error through zero. Increasing exposure helps only while the stream remains observable and the model follows its evolution.

Unknown constant station phases rotate the complex correlation without changing its magnitude, whereas phase fluctuations during averaging reduce coherence and can mimic a larger physical dispersion. Our $\eta_i$ parameters represent constant residual losses. At fixed bulk velocity, removing their calibration priors increases the benchmark transverse error from $0.017$ to $0.021\,\mathrm{m/s}$. Rotation helps separate fixed losses from the changing baseline projections, but the separation depends on network geometry. With the Seattle and Daejeon sites alone, the local fractional width error is $50\%$ with the priors and exceeds unity without them, indicating an unresolved width. Adding Perth reduces the latter to $44\%$, with the bulk velocity fixed in these comparisons. Arbitrary time-dependent losses, correlated noise, or unresolved overlapping streams require a more complete model and can degrade these bounds.

The radiant remains an essential part of the astronomical interpretation. Phase wraps produce competing directional solutions, and a narrow local error does not select the correct fringe. At $f=f_c$, the carrier-phase contribution from the spin-axis component of the bulk velocity is constant on each baseline and can be absorbed into station phases. The visibility and linewidth can still depend on that component through the flow direction and speed, while frequencies away from $f_c$ provide additional phase information. Allowing the speed to vary in a local fit does not establish a unique global velocity. Each viable flow solution must be tested before assigning dispersions to a unique radiant.

The measured dispersions reveal the stream's internal motion along and across the observed bulk flow. Their ratio probes velocity anisotropy within the axisymmetric model of Eq.~\eqref{eq:tensor}. With enough information to fit additional parameters, a general velocity covariance could identify the principal spreads and their orientations. These can be compared with the Galactic flow direction after correcting the mean velocity for the observer's motion. Longitudinal--transverse correlations can also make the mean arrival direction depend on frequency. Tidal shear and phase mixing shape these quantities in ways that depend on the orbit and Galactic potential~\cite{Helmi:1999ks,Tremaine:1998nk}. An old stream need not have a universal flattening or a fixed alignment with its bulk velocity, so the measured covariance tests specific dynamical predictions.

Evolving a minicluster's initial profile and velocity distribution through tidal stripping along a Galactic orbit predicts the present covariance, density, and encounter profile. Joint comparisons constrain stripping history, initial internal motion, and progenitor mass within a specified structural model~\cite{Kavanagh:2020gcy,Visinelli:2026rbm}. Dispersion alone does not uniquely determine age or mass. Population models connect these inferences to the initial small-scale structure predicted by axion production scenarios~\cite{Hogan:1988mp,Eggemeier:2019khm}.

Combined with the stream mass density $\rho_s$, the measured velocity covariance constrains the local phase-space density within the Gaussian model, although unresolved substructure limits this interpretation. Inferring $\rho_s$ from the signal power also requires the axion-photon coupling $g_{a\gamma\gamma}$ and calibrated receiver response, since conversion power scales as $g_{a\gamma\gamma}^2\rho_s$~\cite{Sikivie:2020zpn}. Spectral information on the gravitational potential can separate density and coupling during suitable bound minicluster encounters~\cite{Dandoy2024}, but stream dispersions alone do not.

When a tidal envelope varies slowly compared with travel across a terrestrial baseline, its station-to-station timing carries little directional information. Sharp features could constrain crossing trajectories, although density boundaries need not follow the particle flow. Timing is central for emitted pulses~\cite{Fairhurst:2009tc}, including collapse of axion stars or axion clouds around rotating black holes~\cite{Levkov:2016rkk,Eby,Yoshino:2012kn}. Passing clumps require appropriate density profiles, and relativistic bursts require the relevant dispersion relation and receiver response.

Axion astronomy can ask more than where a signal comes from. Spectroscopy measures the spread of speeds, while spatial coherence exposes the spread of directions. For tidal streams, their combination can recover internal velocity anisotropy that the leading single-receiver spectrum cannot constrain. This gives a global detector network a concrete astronomical task of measuring the dynamical structure of dark matter flowing through the Solar System.

\begin{acknowledgments}
\noindent\textit{Acknowledgments.} We thank Sungwoo Youn for helpful discussions. This work was supported by the U.S. Department of Energy, Office of Science, Office of High Energy Physics, under Award Number DE-SC0010107, and by a ChatGPT for Academic Researchers grant from OpenAI. OpenAI ChatGPT, OpenAI Codex, and Anthropic Claude, accessed during manuscript preparation, were used as research-assistance tools for code development, numerical auditing, workflow documentation, and editorial revision. The authors critically reviewed and validated all AI-assisted material and remain fully responsible for the accuracy, integrity, originality, and scientific content of the manuscript.

\end{acknowledgments}

\bibliography{refs}

\end{document}